\documentclass[journal=cm,manuscript=article,layout=singlecolumn]{achemso}
\usepackage{hyperref}

\usepackage[version=3]{mhchem} 

\author{Hung Ba Tran}
\email{tran.h.ag@m.titech.ac.jp}
\affiliation{Laboratory for Materials and Structures, Institute of Innovative Research, Tokyo Institute of Technology, Midori-ku, Yokohama 226-8503, Japan}
\alsoaffiliation{Quemix Inc., Taiyo Life Nihombashi Building, 2-11-2, Nihombashi Chuo-ku, Tokyo 103-0027, Japan}
\alsoaffiliation{Institute of Scientific and Industrial Research, Osaka University, 8-1 Mihogaoka, Ibaraki, Osaka 567-0047, Japan}

\author{Hiroyoshi Momida}
\affiliation{Institute of Scientific and Industrial Research, Osaka University, 8-1 Mihogaoka, Ibaraki, Osaka 567-0047, Japan}

\author{Yu-ichiro Matsushita}
\affiliation{Laboratory for Materials and Structures, Institute of Innovative Research, Tokyo Institute of Technology, Midori-ku, Yokohama 226-8503, Japan}
\alsoaffiliation{Quemix Inc., Taiyo Life Nihombashi Building, 2-11-2, Nihombashi Chuo-ku, Tokyo 103-0027, Japan}

\author{Kazunori Sato}
\affiliation{Divisions of Materials and Manufacturing Science, Graduate School of Engineering, Osaka University, 2-1 Yamada-oka, Suita, Osaka 565-0871, Japan}
\alsoaffiliation{Center for Spintronics Research Network, Osaka University, Toyonaka, Osaka 560-8531, Japan}

\author{Yukihiro Makino}
\affiliation{Daikin Industries, LTD., 1-1 Nishi-Hitotsuya, Settsu, Osaka 566-8585, Japan}

\author{Koun Shirai}
\affiliation{Institute of Scientific and Industrial Research, Osaka University, 8-1 Mihogaoka, Ibaraki, Osaka 567-0047, Japan}

\author{Tamio Oguchi}
\affiliation{Institute of Scientific and Industrial Research, Osaka University, 8-1 Mihogaoka, Ibaraki, Osaka 567-0047, Japan}
\alsoaffiliation{Center for Spintronics Research Network, Osaka University, Toyonaka, Osaka 560-8531, Japan}

\title[An \textsf{achemso} demo]{Effect of magnetocrystalline anisotropy on magnetocaloric properties of AlFe$_{2}$B$_{2}$ compound}

\begin{document}


\begin{abstract}
It is well known that the temperature dependence of the effective magnetocrystalline anisotropy energy obeys the $l(l+1)/2$ power law of magnetization in the Callen-Callen theory. Therefore, according to the Callen-Callen theory, the magnetocrystalline anisotropy energy is assumed to be zero at the critical temperature where the magnetization is approximately zero. This study estimates the temperature dependence of the magnetocrystalline anisotropy energy by integrating the magnetization versus magnetic field ($M$--$H$) curves, and found that the magnetocrystalline anisotropy is still finite even above the Curie temperature in the uniaxial anisotropy, whereas this does not appear in the cubic anisotropy case. The origin is the fast reduction of the anisotropy field, which is the magnetic field required to saturate the magnetization along the hard axis, in the case of cubic anisotropy. Therefore, the magnetization anisotropy and anisotropic magnetic susceptibility, those are the key factors of magnetic anisotropy, could not be established in the case of cubic anisotropy. In addition, the effect of magnetocrystalline anisotropy on magnetocaloric properties, as the difference between the entropy change curves of AlFe$_{2}$B$_{2}$ appears above the Curie temperature, which is in good agreement with a previous experimental study. This is proof of magnetic anisotropy at slightly above Curie temperature.
\end{abstract}



\section{\label{sec:level1}Introduction}

Heat-assisted magnetic recording is an important method for increasing the capacity and energy efficiency of data storage\cite{bb1,bb2}. Magnetic materials with strong magnetocrystalline anisotropy energy (MAE) are required to maintain the stability of small-sized magnetic particles for a long time. The other side is the difficulty in changing the magnetic state using the write head during the writing process. As the magnetocrystalline anisotropy energy depends on temperature, applying heat can decrease the coercivity (magnetic field required to change the magnetic state) of magnetic material\cite{bb1,bb2}. Therefore, the temperature dependence of MAE is a crucial characteristic of magnetic materials to determine the magnetic field needed to change the magnetization. In early theoretical studies, the effective magnetocrystalline anisotropy constant was found to be proportional to a polynomial of magnetization $M^{3}$ and $M^{10}$ in uniaxial and cubic anisotropy, respectively, \cite{bb3}. Subsequently, the $l(l+1)/2$ power law (where $l$ is the order of anisotropy constants) of magnetocrystalline anisotropy energy, the so-called Callen-Callen theory, was reproduced in previous studies by using the constrained Monte-Carlo method\cite{bb4,bb5}. If the magnetization is equal to zero at the Curie temperature, MAE should be equal to zero\cite{bb3,bb4,bb5}. However, finite magnetocrystalline anisotropy at slightly above the Curie temperature has been observed in recent experimental studies\cite{bb6,bb7}.

The application of magnetocrystalline anisotropy is the rotating magnetocaloric effect\cite{bb8,bb9,bb10,bb11,bb12,bb13,bb14}. The temperature and heat exchange of the anisotropic magnetic material can be controlled by rotating the sample (or changing the relative direction of the external magnetic field) at a fixed strength of the applied magnetic field. The origin of this effect is the modification of the magnetocrystalline anisotropy on the magnetization when changing the relative direction of the external magnetic field. The isothermal magnetic entropy change has a peak at the critical temperature because the derivative of magnetization diverges at the ferromagnetic-paramagnetic (FM-PM) transition\cite{bb15,bb16}. If the magnetocrystalline anisotropy disappears at the Curie temperature, the modification of the magnetocrystalline anisotropy only appears on the low-temperature side of the entropy change peak. However, several experimental studies have found that the isothermal magnetic entropy change is modified not only on the low-temperature side but also on the high-temperature side of the main peak when a magnetic field is applied in different directions\cite{bb8,bb9,bb10,bb11,bb12,bb13,bb14}. The magnetocrystalline anisotropy energy affects the isothermal magnetic entropy change even above the Curie temperature in the anisotropic magnetocaloric effect.

\begin{figure}[htp] 
\centering
\includegraphics[width=7.4cm]{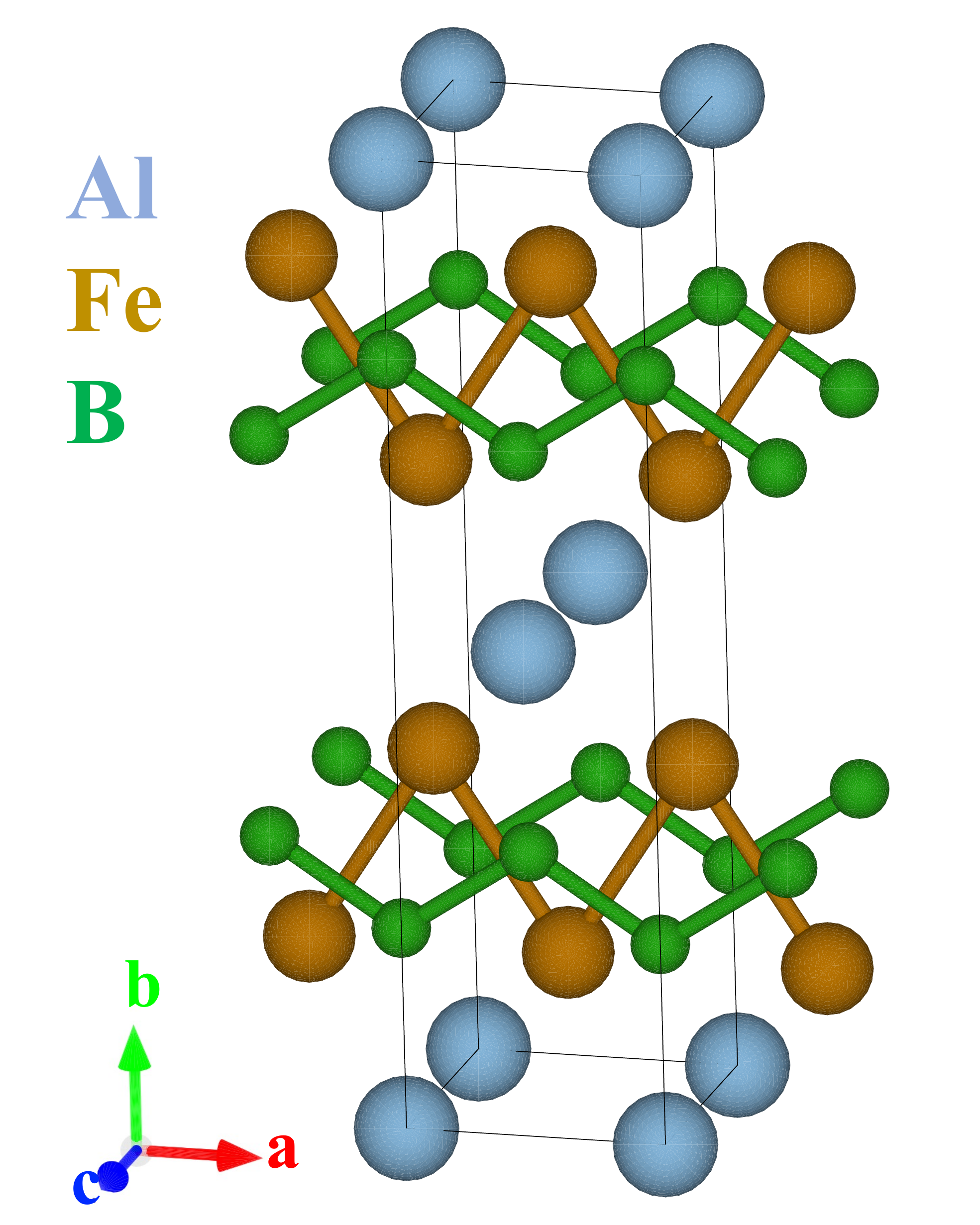} 
\caption{ The crystal structures of orthorhombic AlFe$_{2}$B$_{2}$. The gray, orange, and blue spheres indicate the Al, Fe, and B atoms, respectively.} 
\label{FIG1}
\end{figure}

In recent years, AlFe$_{2}$B$_{2}$ has attracted much attention for future refrigeration with earth-abundant elements, room-temperature magnetic transitions, and strong magnetocrystalline anisotropy energy\cite{bb17,bb18,bb19,bb20,bb21}. AlFe$_{2}$B$_{2}$ has an orthorhombic layered crystal structure with a space group $Cmmm$ as depicted in Figure \ref{FIG1}. The Fe and B layers form a zigzag chain in the $ac$ plane. Furthermore, the Fe and B layers are separated by the Al plane. The Curie temperature of the material is varied in the range of 274--320 K, which depends on the synthesis method \cite{bb18}. AlFe$_{2}$B$_{2}$ has an easy ([100]), intermediate ([010]), and hard ([001]) magnetic axes \cite{bb17}. The magnetocrystalline anisotropy constants (at 2 K) were reported in a previous experimental study with $K_{\rm 001}$=1.8 (MJ/m$^{3}$) and $K_{\rm 010}$=0.23 (MJ/m$^{3}$)\cite{bb17}. The dependence of magnetocaloric properties on the direction of an applied magnetic field has been reported in previous experimental study\cite{bb18}. However, to date, there has been no comprehensive theoretical study on the rotating magnetocaloric properties of the material.

In the present work, we first estimate the temperature dependence of the MAE in the uniaxial and cubic anisotropy cases by considering the energy cost to magnetize in some specific directions in simple-cubic lattice model. Moreover, as MAE is estimated as the area between two magnetization curves when applying magnetic fields along the hard and easy axes, the ordinary Monte-Carlo method is considered. The magnitude of the projected magnetization can be saturated when a the magnetic field is applied along the hard axis by increasing the strength of the external magnetic field. The results are compared with the well-known Callen-Callen theory for the uniaxial and cubic anisotropy cases. In addition, the temperature dependence of the magnetocrystalline anisotropy constant and magnetocaloric effect in orthorhombic AlFe$_{2}$B$_{2}$ is also considered. This material has easy, intermediate, and hard axes, which can exhibit different behavior in the temperature dependence of magnetic anisotropy constants. In addition, the entropy change at a magnetic field of 2 T is compared with an experimental work\cite{bb18}.

\section{\label{sec:level2}Methodologies}
The electronic structure calculations and structure optimizations of orthorhombic AlFe$_{2}$B$_{2}$ were performed using DFT using the full-potential linearized augmented plane wave (FLAPW) method\cite{bb22}. The lattice constants ($a=2.9168 \AA$, $b=11.033 \AA$, and $c= 2.8660 \AA$) were used for AlFe$_{2}$B$_{2}$\cite{bb17}. The atomic position relaxation was carried out by optimizing all atomic forces of less than 1 mRy/Bohr. The generalized gradient approximation (GGA) was used for exchange and correlation\cite{bb23}. Magnetocrystalline anisotropy energy is evaluated as the energy necessary to rotate the quantization axis with spin orbit coupling (SOC) in the second variation steps\cite{bb24}. The convergence of magnetocrystalline anisotropy energy was achieved using a large number of \textit{k}-point meshes of 48$\times$16$\times$48.   

The magnetic exchange constants ($J_{ij}$) of stoichiometric AlFe$_{2}$B$_{2}$ were evaluated using the Liechtenstein formula in the Korringa-Kohn-Rostoker (KKR) method with coherent potential approximation (CPA)\cite{bb25,bb26,bb27}:

\begin{equation}
J_{ij}=\frac{1}{4\pi}\mathrm{Im}\int^{E_F}dE\mathrm{Tr}\left \{ \Delta t_i(E)T_{ij}^\uparrow\Delta t_j(E)T_{ij}^\downarrow \right \},
\label{Eq1}
\end{equation}

\noindent where $T_{ij}^\uparrow$ and $T_{ij}^\downarrow$ are off-diagonal scattering path operators for spin-up and spin-down between atomic sites $i$ and $j$, respectively, and $\Delta t_i(E)=t_{i}^{\uparrow}-t_{i}^{\downarrow}$ with the single site $\textit{t}$-matrix $t_{i}^{\sigma}$ for spin $\sigma$ at site \textit{i}.

The Heisenberg model with uniaxial and cubic anisotropy terms was used to obtain the magnetism and magnetocaloric properties at a finite temperature\cite{bb4}:

\begin{equation}
\begin{split}
&H_{\rm Heis}=-\sum_{<ij>}J_{ij}^{m}\overrightarrow{S_{i}}\overrightarrow{S_{j}}-\sum_{i}k_{\rm u}(\overrightarrow{e_{u}}\overrightarrow{S_{i}})^{2}\\
&-\sum_{i}k_{\rm c}(S_{ix}^{4}+S_{iy}^{4}+S_{iz}^{4})-g\mu _{\rm B}\sum_{i}\overrightarrow{H_{\rm ext}}\overrightarrow{S_{i}},
\end{split}
\label{Eq2}
\end{equation}  
 
\noindent where $g$ is the $g$-factor, $\mu_{\rm B}$ is the Bohr magneton, $k_{\rm u}$ is the uniaxial anisotropy constant, and $k_{\rm c}$ is the cubic anisotropy constant. The first term expresses the exchange interactions between spins at sites \textit{i} and \textit{j}. The spin tends to be parallel to the neighboring site when the magnetic exchange coupling constant $J_{ij}^{m}$ is positive. The spin is antiparallel to the neighboring spin for negative $J_{ij}^{m}$. The second term in Eq. (\ref{Eq2}) is the interaction between the spin at site \textit{i} and the uniaxial anisotropy, $e_{\rm u}$ is the direction of the easy axis in the case of a positive $k_{\rm u}$. The third term in Eq. (\ref{Eq2}) is the cubic anisotropy term with $S_{ix}$ as the $x$ component of $S_{i}$. The final term in Eq. (\ref{Eq2}) is the interaction of the spin at site $i$ with the external magnetic field. At a finite external magnetic field, the spin tends to be parallel to the direction of the magnetic field to minimize the energy. The size of the simple-cubic lattice is assumed to be 80$\times$80$\times$80 unit cells, and the first nearest neighbor of the exchange interaction is considered. However, the size of the Monte-Carlo simulation box in the AlFe$_{2}$B$_{2}$ compound is 30$\times$30$\times$30 unit cells with 270000 atoms. The number of Monte-Carlo steps is 200000, and the first 100000 steps are discarded. The Metropolis algorithm was used to achieve thermal equilibrium in the Heisenberg model.

The magnetocrystalline anisotropy energy was estimated by integrating the magnetization versus the magnetic field ($M$--$H$) curves:

\begin{equation}
{\rm MAE}=\int_{0}^{H_{\rm ext}} \{ M(H\parallel{\rm easy\;axis})-M(H\parallel{\rm hard\;axis})\}dH
\label{Eq3}
\end{equation}

\noindent where the magnetic field is used to obtain the saturation of magnetization. The temperature dependence of magnetocrystalline anisotropy energy can be estimated using integration because the area between the two magnetization curves can be measured as the energy cost to magnetize in a certain direction.

The magnetic susceptibility is calculated as

\begin{equation}
\chi = \frac{\partial M}{\partial H} = \frac{(\left \langle M^{2} \right \rangle -\left \langle M \right \rangle^{2})}{k_{\rm B} T},
\label{Eq4}
\end{equation}

\noindent where $k_{\rm B}$ is Boltzmann constant. The magnetic susceptibility is the fluctuation of magnetization in the Monte-Carlo simulation.

The isothermal magnetic entropy changes ($\Delta S_{\rm M}$) are estimated by integrating the derivative of the magnetization ($M$) based on the Maxwell relation\cite{bb15,bb16}

\begin{equation}
\begin{split}
& \Delta S_{\rm M}(H_{\rm ext},T) =\int_{0}^{H_{\rm ext}}\left (\frac{\partial M(H,T) }{\partial T} \right ), \\
& \cong \sum_{j=0}^{N}\frac{M(H_j,T+\Delta T)-M(H_j,T-\Delta T)}{2\Delta T}\Delta H
\end{split}
\label{Eq5}
\end{equation}

\noindent The isothermal magnetic entropy change is obtained by integrating with fine mesh in temperature and external magnetic field. $\Delta T$ and $\Delta H$ are taken as 2 K and 0.2 T, respectively. Owing to the magnetocrystalline anisotropy energy, the isothermal magnetic entropy changes strongly depend on the direction of the applied magnetic field. 

\section{\label{sec:level3}Results and discussion}

\subsection{Uniaxial and cubic anisotropies in simple-cubic lattice model}

\begin{figure}  
\centering
\includegraphics[width=8.4cm]{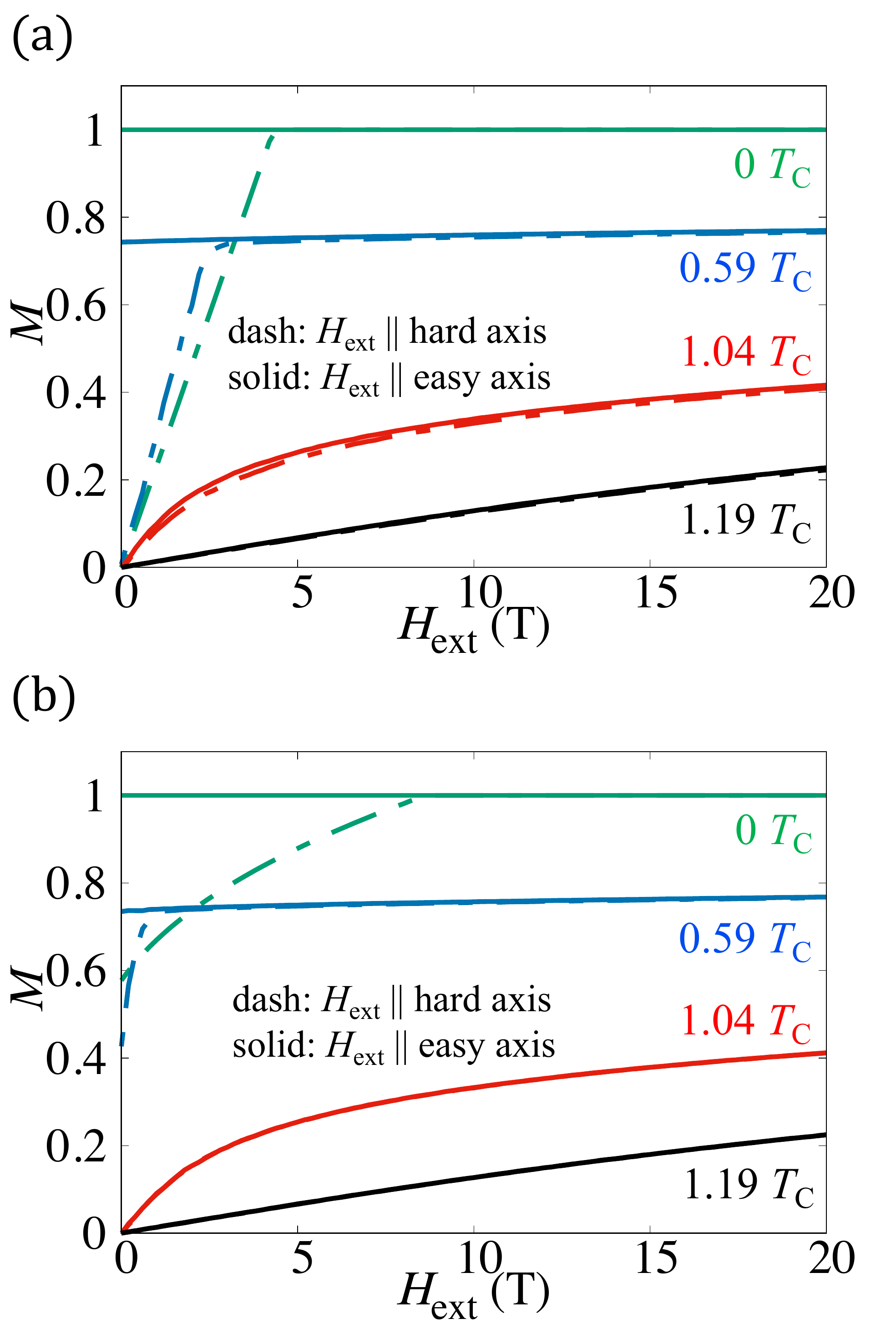} 
\caption{Magnetization-magnetic fields ($M$--$H$) curve at several temperatures when applying magnetic field along the hard and easy axis of uniaxial anisotropy (a) and cubic anisotropy (b) of simple cubic lattice model. The magnetization was normalized to 1.} 
\label{FIG2}
\end{figure}

As magnetic materials usually exhibit uniaxial or cubic anisotropy, a simple-cubic lattice with uniaxial and cubic anisotropy interaction was considered in the first part to provide a general characterization of uniaxial and cubic anisotropy. The magnetic exchange coupling constant of the first nearest neighbor, equal to 10 meV, was used in this simulation. The $k_{\rm u}$ and $k_{\rm c}$ were 0.25 meV, and the size of the magnetic moment was 2.0 $\mu_{\rm B}$. The magnetization-magnetic field ($M$--$H$) curves at several temperatures in the uniaxial anisotropy case are shown in Figure \ref{FIG2} (a). At 0 K, increasing the magnetic field along the hard axis leads to a linear enhancement of magnetization before saturating at $H_{\rm A}= 2 K _{\rm u}/M_{\rm S}$ as the Stoner-Wohlfarth model for magnetic reversal. When the temperature is increased, the saturation magnetization at zero external magnetic field is decreased. Furthermore, the anisotropy field, which is the required magnetic field to saturate the magnetization along the hard axis, is also decreased. This results to a high order of dependence of MAE on magnetization rather than linear dependence. At an intermediate temperature, the magnetization curve when the magnetic field along the hard axis has a wing when the magnetic field approaches the anisotropy field. This wing is related to the finite temperature effect on the direction of magnetization, and the direction of magnetization is difficult along the hard axis. Furthermore, a high external magnetic field is required to approach the saturation magnetization when a magnetic field is applied along the hard axis at these temperatures. If the magnetization curve of the hard axis is extrapolated to the limit of zero, by applying external magnetic field as the saturation magnetization of the hard axis, there is a finite value of the difference in saturation magnetization between the easy and hard axes, the so-called magnetization anisotropy. At a temperature slightly above the Curie temperature, the magnetization curve indicates when the magnetic field along the hard axis is not saturated. This means that magnetizing along the hard axis in the uniaxial case is still harder than the magnetization along the easy axis even above the Curie temperature, which was also observed in the experimental study\cite{bb6}. This is due to the difference in the slope of the $M$--$H$ curves in the easy and hard axes at a finite magnetic field applied, such as the anisotropic magnetic susceptibility. The magnetic susceptibility is the fluctuation of the magnetization curves at a fixed strength of the external magnetic field. Therefore, the dependence of magnetic susceptibility on the direction of the external magnetic field at these temperatures implies that the external magnetic field leads to the enhancement of the anisotropic magnetic susceptibility. When the magnetization is enhanced by the applied field at a temperature slightly higher than the Curie temperature, the effect of magnetic anisotropy appears and enhances the first derivative of magnetization in the field, as the magnetic susceptibility.   

The $M$--$H$ curves at several temperatures in the cubic anisotropy case are shown in Figure \ref{FIG2} (b). At 0 K, the magnetization along the hard axis at zero external magnetic fields is non-zero because the hard and easy axes are not perpendicular in the cubic case. Moreover, it is understood that the anisotropy field is higher in this case than that in the case of uniaxial anisotropy. However, the anisotropy field is the same as 0.25 meV in both the cases. If the temperature increases, the magnetization and anisotropy fields also decrease as the temperature affects the order parameter and MAE. However, the anisotropy field is decreased rapidly compared with the uniaxial case. The cubic anisotropy is a higher order of anisotropy interaction than uniaxial anisotropy, as shown in Eq. (\ref{Eq2}). At high temperatures, the area between the two magnetization curves is diminishes to almost zero, in contrast to the uniaxial case. This means that the magnetocrystalline anisotropy energy above the Curie temperature is approximately equal to zero. Because the magnetocrystalline anisotropy energy is related to the energy barrier in free energy, which contains the energy and entropy terms of cubic and uniaxial anisotropy, different at temperatures above the Curie temperature. The high external magnetic field cannot enhance the anisotropic magnetic susceptibility compared with the uniaxial case, as discussed above. One possibility for this phenomenon is the shape of the magnetic anisotropy energy surface of cubic anisotropy, which depends on the direction of magnetization, is isotropic and does not depend on the strength of the external magnetic field. 

\begin{figure}  
\centering
\includegraphics[width=8.4cm]{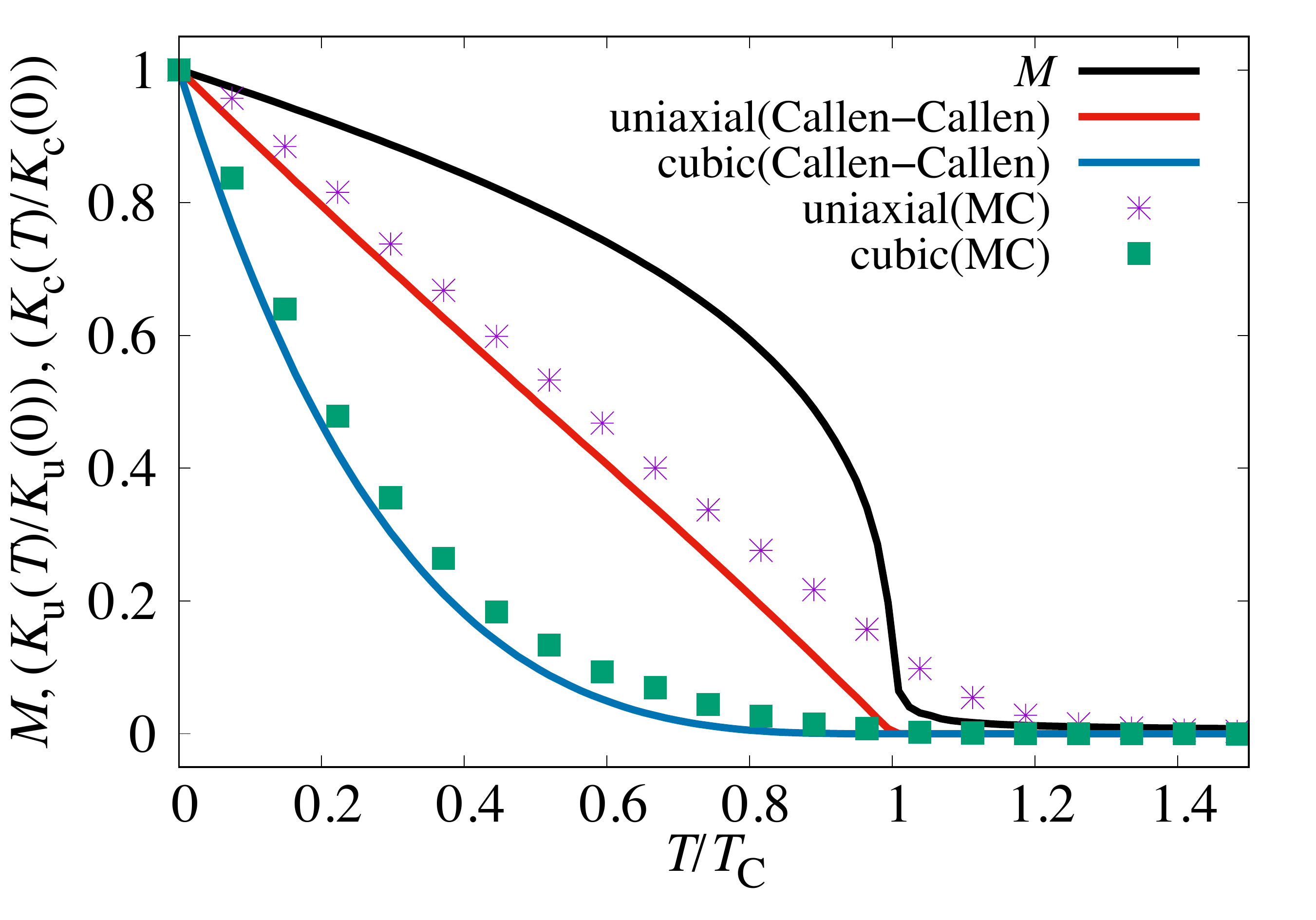} 
\caption{Temperature dependence of magnetization (black line), anisotropy constant of uniaxial (red line) and cubic (blue line) anisotropies of Callen-Callen model as $M^{3}$ and $M^{10}$, respectively. The integration of $M$--$H$ curves for uniaxial (crossed purple points) and cubic (square green points) anisotropies using the Monte-Carlo method.} 
\label{FIG3}
\end{figure}

The temperature dependence of magnetization and normalized anisotropy constant of uniaxial and cubic anisotropies in a simple-cubic lattice model are shown in Figure \ref{FIG3}. The magnetization curve at zero external magnetic fields was used to determine the effective anisotropy constant of the Callen-Callen theory. The results of the integration $M$--$H$ curve of cubic anisotropy are in good agreement with $M^{10}$, whereas the critical temperature of the uniaxial anisotropy constant is higher than the Curie temperature defined by the Callen-Callen theory. The uniaxial and cubic anisotropy constants of the Callen-Callen theory are obtained from the power law of magnetization. This means that when the magnetization becomes approximately equal to zero at the Curie temperature, the effective anisotropy constants are also assumed to be zero. The different behavior between uniaxial and cubic anisotropy may be due to the shape of the magnetic anisotropic energy surface, which is the energy that depends on the direction of magnetization with a finite external magnetic field. In uniaxial anisotropy, the external magnetic field enhances the anisotropic magnetic susceptibility near the Curie temperature. It is preferred to magnetize along the easy axis compared to the other axes. Therefore, the magnetization along the hard axis, which is the integration of magnetic susceptibility, is slightly smaller than along the easy axis above the Curie temperature. On the other hand, the anisotropic magnetic susceptibility of cubic anisotropy above the Curie temperature becomes approximately zero. Moreover, the combination of cubic anisotropy and the external magnetic field is almost unchanged when magnetizing along the hard and easy axes. This phenomenon leads to different behaviors in the magnetocaloric effect of uniaxial and cubic anisotropy because the cubic anisotropy energy disappears above the Curie temperature, while the uniaxial anisotropy energy is not negligible at these temperatures.  

We derived the relation of magnetization anisotropy and anisotropic magnetic susceptibility with MAE from the definition of MAE and the magnetic susceptibility of Eq. (\ref{Eq3}) and Eq. (\ref{Eq4}) as

\begin{equation}
\begin{split}
&\frac {\partial E_{\rm MAE}}{\partial H}\\
&=\frac {\partial F(H_{\rm ext}\parallel{\rm easy},T)}{\partial H} - \frac {\partial F(H_{\rm ext}\parallel{\rm hard},T)}{\partial H}, \\
&=M(H_{\rm ext}\parallel{\rm easy},T)-M(H_{\rm ext}\parallel{\rm hard},T)\\
\end{split}
\label{Eq6}
\end{equation}

\begin{equation}
\begin{split}
&\frac {\partial^{2} E_{\rm MAE}}{\partial H^{2}}\\
&=\frac {\partial^{2} F(H_{\rm ext}\parallel{\rm easy},T)}{\partial H^{2}} - \frac {\partial^{2} F(H_{\rm ext}\parallel{\rm hard},T)}{\partial H^{2}}, \\
&=\chi(H_{\rm ext}\parallel{\rm easy},T)-\chi(H_{\rm ext}\parallel{\rm hard},T)\\
\end{split}
\label{Eq7}
\end{equation}

From the Eqs. (\ref{Eq6}) and (\ref{Eq7}), magnetization anisotropy is the first derivative while the anisotropic magnetic susceptibility is the second derivative of MAE on the magnetic field. This means that their effects on MAE can be considered in our Monte-Carlo simulation as the integration of $M$--$H$ curves. The magnetization anisotropy and anisotropic magnetic susceptibility lead to finite MAE of the uniaxial case and negligible MAE of the cubic case at a temperature slightly higher than the Curie temperature. Because the magnetization anisotropy is the integration of anisotropic magnetic susceptibility until the anisotropy field is reached, it only appears below the Curie temperature where the anisotropy field is finite. Moreover, because the anisotropy field of the cubic anisotropy decreases much faster than the uniaxial anisotropy, the magnetization anisotropy may not appear in the cubic case.

\subsection{Pristine AlFe$_{2}$B$_{2}$}

\begin{figure}[ht]  
\centering
\includegraphics[width=8.4cm]{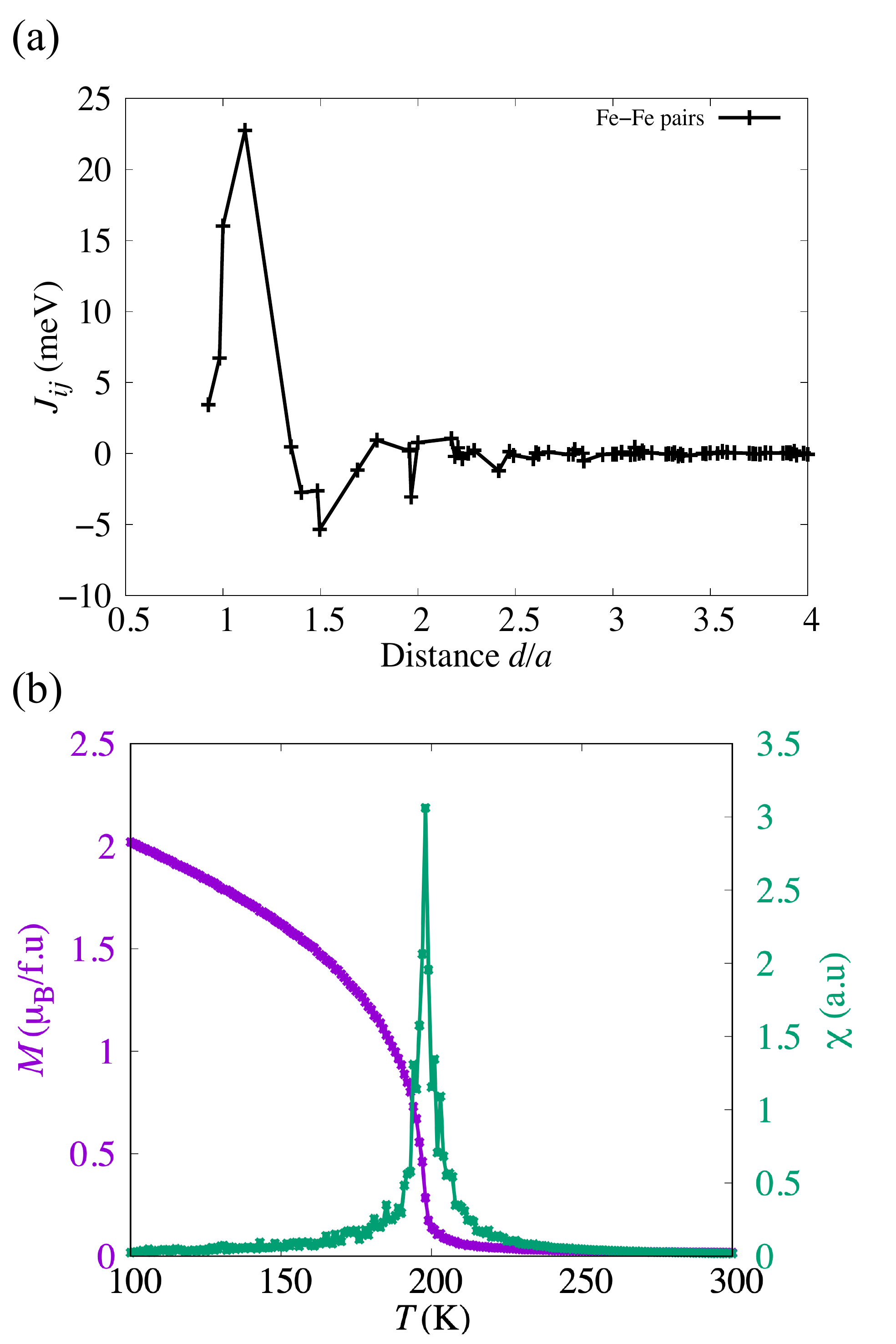} 
\caption{(a) Magnetic exchange coupling constants of the Fe-Fe pairs as a function of Fe-Fe distance ($d/a$) normalized with the lattice constant $a$. (b) Magnetization (purple line) and magnetic susceptibility (green line) curves of the Monte-Carlo simulation at zero external magnetic field.} 
\label{FIG4}
\end{figure}

The magnetic exchange coupling constants between Fe atoms of orthorhombic AlFe$_{2}$B$_{2}$ are shown in Figure \ref{FIG4} (a). The coupling constants exhibit a long-range behavior with oscillation when the distance increases. The magnetic exchange coupling constants from the first to fourth nearest neighbors are positive. Furthermore, the coupling constants are increased when the distance is increased. This is not an ordinary direct-exchange type of magnetic interaction. The first, second, and third nearest neighbors are surrounded by B atoms inside the Fe-B layer. Therefore, the $p$-state of B affects the coupling constants via a super-exchange interaction, and the coupling constants increase when the distance increases in this case. On the other hand, the fourth nearest neighbor is a pair of Fe atoms in different layers separated by the Al plane. It leads to the highest value of the coupling constant of the fourth nearest neighbor as direct exchange interaction. The Curie temperature can be estimated from the magnetic exchange coupling constants using the mean-field approximation (MFA) and Monte-Carlo method. The Curie temperature obtained by MFA is 365 K, while the value of the Monte-Carlo method is 199 K. The Curie temperature of the MFA is usually overestimated, whereas the Monte-Carlo method should be more precise. The magnetization and magnetic susceptibility of AlFe$_{2}$B$_{2}$ are shown in Figure \ref{FIG4} (b). The Curie temperature of the Monte-Carlo method is obtained as the divergence of the magnetic susceptibility.

\begin{figure*}[ht]  
\centering
\includegraphics[width=16.0cm]{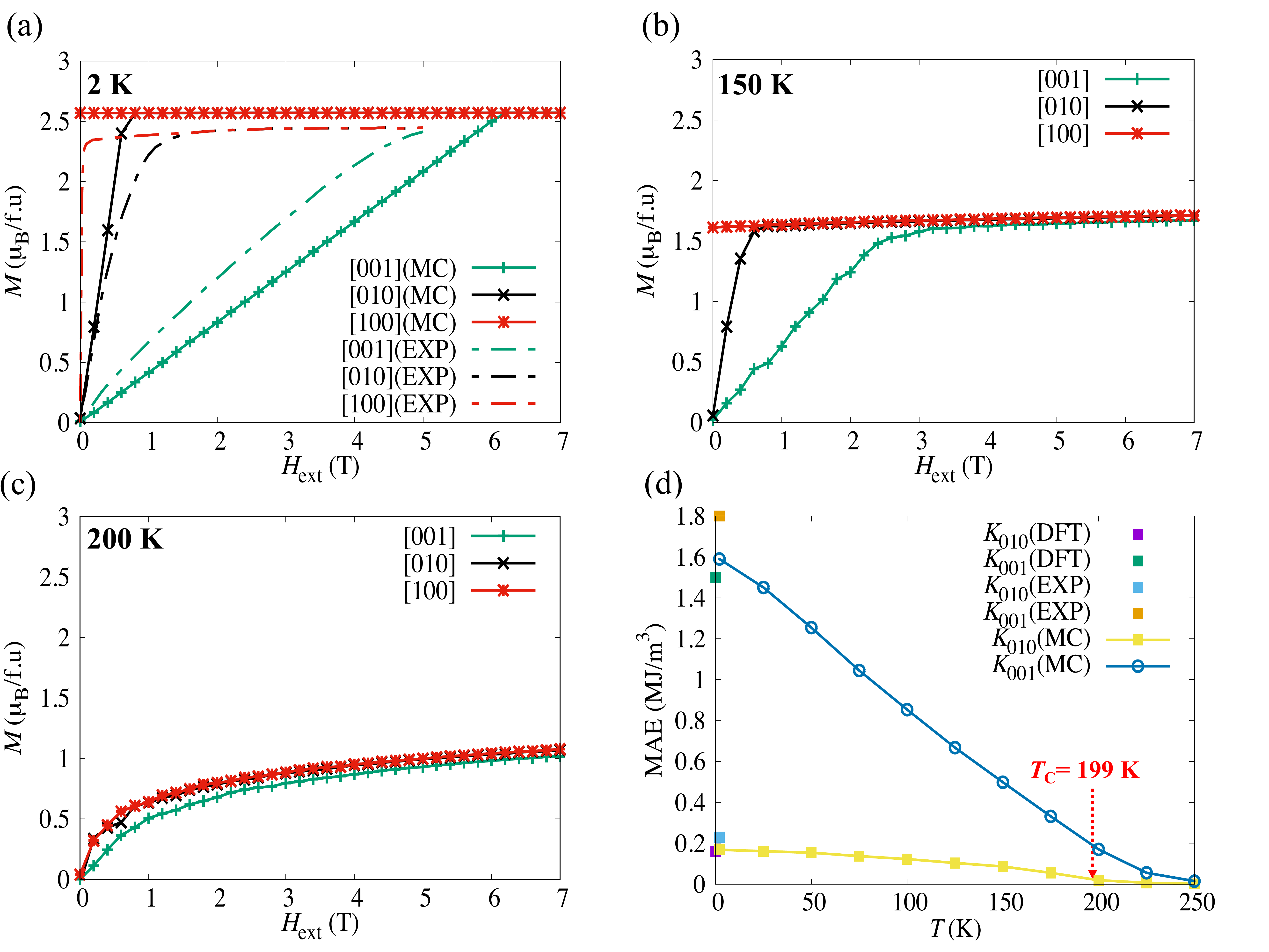} 
\caption{ (a), (b), (c) Magnetization-magnetic field ($M$--$H$) curves of several temperature of Monte-Carlo method when magnetic field along the easy ($[100]$), intermediate ($[010]$), and hard ($[001]$) axis. The $M$--$H$ curves at 2 K in the experimental study \cite{bb17} are shown as dashed lines. (d) Temperature dependence effective magnetocrystalline anisotropy energy of the Monte-Carlo method and available data from first-principles calculations (0 K) and experimental studies (2 K)\cite{bb17}. } 
\label{FIG5}
\end{figure*}
 
The magnetization-magnetic field $M$--$H$ curves of AlFe$_{2}$B$_{2}$ at several temperatures are shown in Figure \ref{FIG5} (a), (b), and (c). The magnetization curves obtained when applying a magnetic field along the hard, intermediate, and easy axes at 2 K in the Monte-Carlo simulation (the solid line) are in good agreement with the experimental results (the dash line)\cite{bb17}. The magnetocrystalline anisotropy energy can be estimated as the area between two magnetization curves, one for the easy axis and the other for the hard or intermediate axis. Applying an external magnetic field along the hard axis requires a large magnetic field to saturate the magnetization. The anisotropy field in the case of the hard axis of the Monte-Carlo method is slightly larger than the experimental results, whereas the anisotropy field of the intermediate axis is slightly smaller than the experimental value. When the temperature is increased, the magnetization and anisotropy field are decreased. However, below the Curie temperature, the anisotropy field of the hard axis is decreased much faster than that in the intermediate case. At 200 K (slightly above the Curie temperature of 199 K), the magnetization is approximately zero at zero external magnetic fields. However, applying an external magnetic field enhances the area between the magnetization curves at this temperature. The area between the magnetization curves of the easy and intermediate axes is approximately equal to zero, whereas it is not negligible between the easy and hard axes. This means that the magnetocrystalline anisotropy constant of the intermediate axis is approximately equal to zero, whereas the hard axis is not negligible at these temperatures. 
 
The magnetocrystalline anisotropy energy constants of AlFe$_{2}$B$_{2}$ are shown in Figure \ref{FIG5} (d). The anisotropy energy obtained using the present first-principles DFT calculations of the hard axis is 1.5 MJ/m$^{3}$, which is close to the experimental result of 1.8 MJ/m$^{3}$ at 2 K\cite{bb17}. Besides, the value of the intermediate case of DFT is 0.16 MJ/m$^{3}$, which slightly smaller than the value of experimental study 0.23 MJ/m$^{3}$\cite{bb17}. By integrating the $M$--$H$ curves in the Monte-Carlo method, we can consider the temperature dependence of the magnetocrystalline anisotropy energy between the hard and intermediate axes with the easy axis. The anisotropy constant of the hard axis is decreased almost linearly as the temperature is increased. Moreover, the uniaxial anisotropy in the previous section is still finite above the Curie temperature. On the other hand, the magnetocrystalline anisotropy constant of the intermediate axis is decreased slowly below the Curie temperature. Near the Curie temperature, the anisotropy constant decreased rapidly. Furthermore, it became approximately equal to zero at the Curie temperature. This means that the energy barrier between the easy and intermediate axes disappears at the Curie temperature.

\begin{figure*}  
\centering
\includegraphics[width=16.0cm]{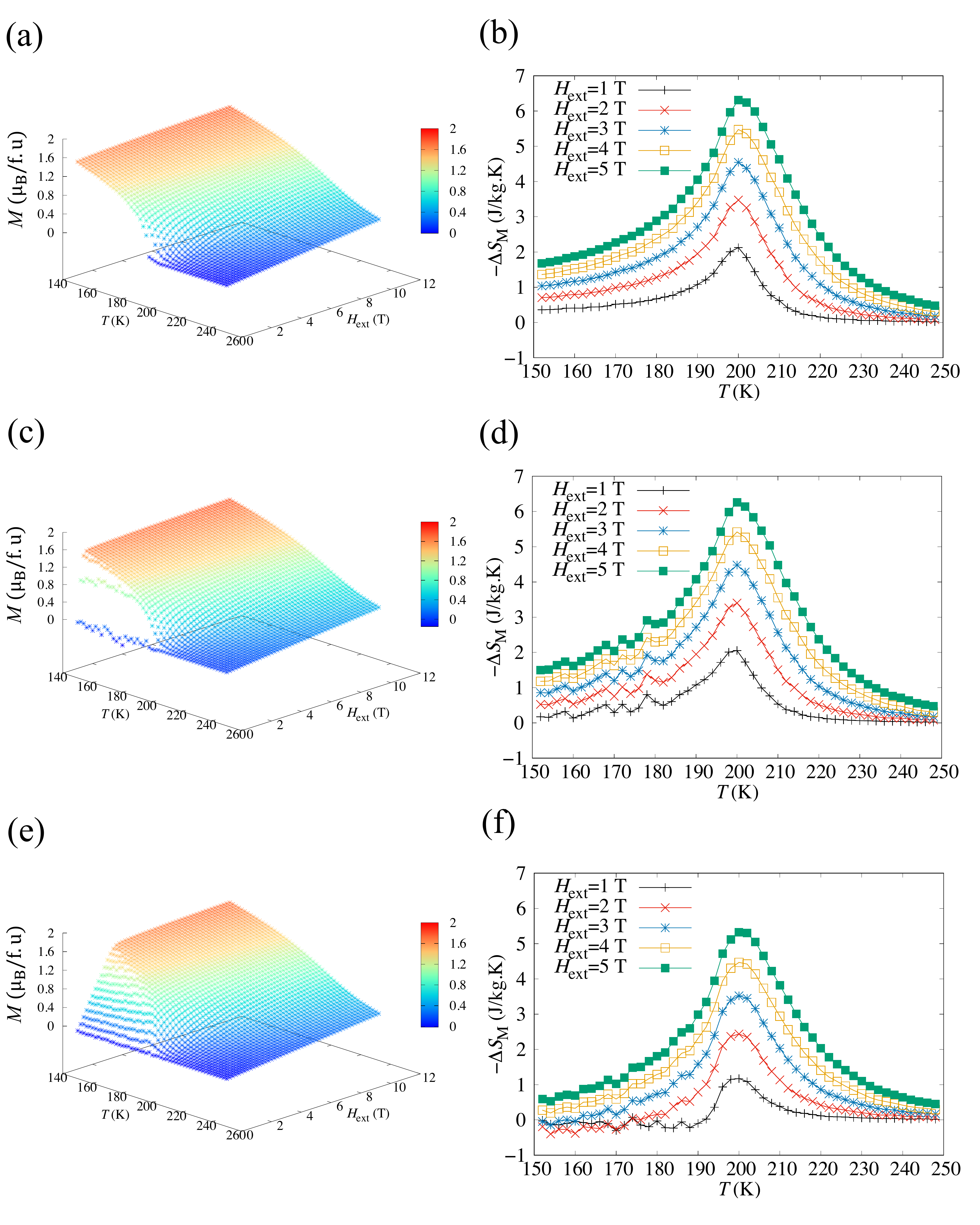} 
\caption{Isothermal magnetization curves as a function of temperature and external magnetic fields ($H_{\rm ext}$) along (a) easy ([100]), (b) intermediate ([010]), and (c) hard ([001]) axes. The isothermal magnetic entropy change ($\Delta S_{\rm M}$) curves as a function of temperature for $H_{\rm ext}$ along the (d) easy ([100]), (e) intermediate ([010]), and (f) hard ([001]) axes.} 
\label{FIG6}
\end{figure*}

The isothermal magnetization and entropy change curves are shown in the Figure \ref{FIG6}. When a magnetic field is applied along the easy axis ([100]), the magnetization decreases when the temperature increases and becomes zero at the Curie temperature as the limit of the magnetic field being zero. By increasing the magnetic field, the magnetization is enhanced, and the transition temperature is shifted. From the magnetization curves, the isothermal magnetic entropy change can be estimated using Maxwell’s relation\cite{bb15,bb16}. The entropy change curves have a peak at the Curie temperature, where the derivative of the magnetization diverges. The value of the peak is increased with increasing strength of the external magnetic field. On the other hand, the magnetic anisotropy became dominant in the low-magnetic-field and low-temperature region in the case of the intermediate ([010]) and hard ([001]) axes. However, the difference in magnetization with the easy axis is decreased by increasing the temperature or magnetic field . At a sufficient magnetic field at finite temperatures, the magnetization looks similar to that in the easy axis case. Therefore, the magnetocrystalline anisotropy energy leads to an extensive modification of the entropy change at a low external magnetic field. The difference in the entropy change curves remain constant at a sufficient magnetic field. There is noise in the magnetization in the temperature range near the Curie temperature due to the competition between magnetic anisotropy and the external magnetic field, which tends to force the magnetization along its own direction. This noise disappears at a sufficiently large magnetic field and high temperature. As a result of this noise in magnetization, the $\Delta S_{\rm M}$ curves of the intermediate and hard axes become noisy in the temperature range. Moreover, owing to the effect of MAE, the entropy change of the hard axis was the lowest at the magnetic field of 1 T. The difference in entropy change remain at a higher external magnetic field since the entropy change is the integration of magnetization curves.

\begin{figure}  
\centering
\includegraphics[width=8.4cm]{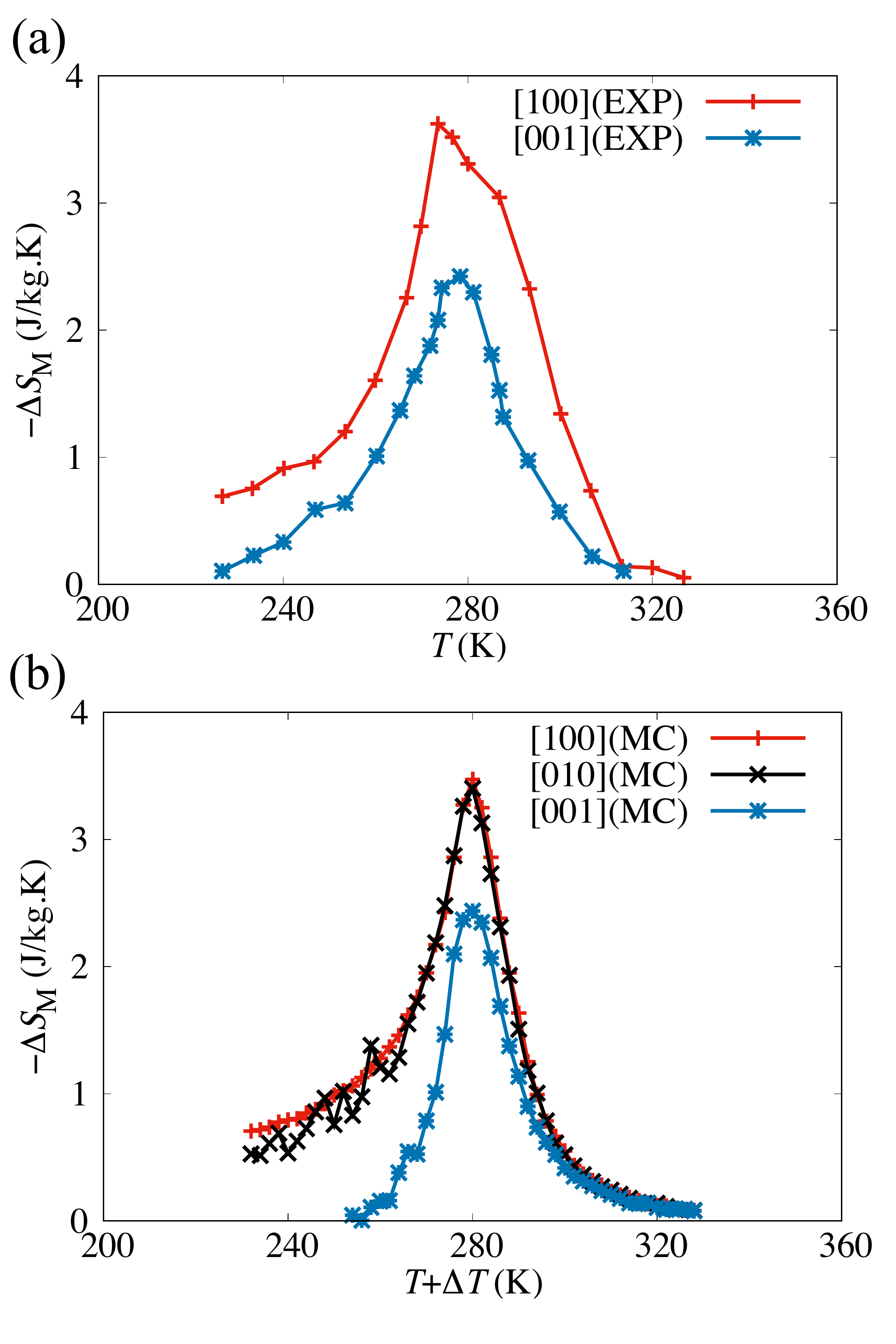} 
\caption{Isothermal magnetic entropy change when magnetic field equal 2 T and along different axis of (a) experimental study \cite{bb18} and (b) Monte-Carlo simulations. $\Delta T$ is used as 80 K to shift the entropy change peak for a simple comparison. } 
\label{FIG7}
\end{figure}

The isothermal magnetic entropy change when the magnetic field is 2 T in an experimental study \cite{bb18} and Monte-Carlo simulation are shown as Figure \ref{FIG7}. $\Delta S_{\rm M}$ calculated with the Monte-Carlo method is in semi-quantitatively good agreement with the experimental results. The entropy change when applying a magnetic field along the easy axis is the highest because the magnetization along the easy axis is assisted by the magnetic anisotropy. The entropy change peak of the experimental results is wider than our result because the Curie temperature by the Monte-Carlo method is smaller than the experimental value. In addition, the discrepancy between the two entropy change curves is due to the modification of the magnetic anisotropy on the magnetization curve. The gap between the entropy change curves of the hard and easy axes is decreased when the temperature increased and disappears slightly above the Curie temperature in both experimental and theoretical studies. This means that the MAE is decreased when the temperature increased and finally disappears slightly above the Curie temperature. Because the anisotropy constant of the intermediate axis is negligible above the Curie temperature, the difference in entropy change of the intermediate and the easy axis is minor. Therefore, there is a nearly easy plane and hard axis at the Curie temperature owing to the temperature dependence of the magnetic anisotropy.

\section{\label{sec:level4}Conclusion}
The temperature dependence of the uniaxial and cubic anisotropy constants was considered using Monte-Carlo simulations for the classical Heisenberg model. The temperature dependence of the cubic anisotropy constant is found to be in good agreement with the Callen-Callen theory, while the critical temperature of the uniaxial anisotropy constant is slightly higher than the Curie temperature. In addition, the magnetism and magnetocaloric properties of orthorhombic AlFe$_{2}$B$_{2}$ were reasonably predicted. The temperature dependence of the magnetocrystalline anisotropy energy of the hard $K_{001}$ and intermediate $K_{010}$ cases demonstrates different behaviors. The good agreement in isothermal magnetic entropy change emphasizes that the magnetic anisotropy affects the magnetization curves even above the Curie temperature. 


\section*{Acknowledgements}
The authors thank Tetsuya Fukushima for his help with first-principles calculations. This work was partly supported by the Di-CHiLD project (Grant No. J205101520). The computation in this work was performed using the facilities of the Supercomputer Center, Institute for Solid State Physics, University of Tokyo and the Cybermedia Center of Osaka University.


\bibliography{basename of .bib file}

\end{document}